# Laser-induced metal halide perovskite-rGO nanoconjugates as anode material in Zn-air supercapacitors

Athanasia Kostopoulou,*[,a] Dimitra Vernardou, *[,b] Nikolaos Livakas, [a] Konstantinos Brintakis,[a] Stelios Daskalakis,[b] and Emmanuel Stratakis*[,c,d]

* Corresponding authors

[a]Institute of Electronic Structure and Laser, Foundation for Research and Technology - Hellas, Heraklion, 71110 Crete, Greece. E-mails: akosto@iesl.forth.gr, stratak@iesl.forth.gr.

[b]Department of Electrical & Computer Engineering, School of Engineering, Hellenic Mediterranean University, Heraklion,710 04 Crete, Greece. Email: dvernardou@hmu.gr.

[c]Department of Physics, University of Crete, 71003 Heraklion, Greece.

Electronic Supplementary Information (ESI) available

**This work reveals that the direct conjugation of metal halide perovskite nanocrystals on rGO sheets can provide high performance and stable electrodes for Zn-air supercapacitors. In particular, we provide a simple, rapid and room temperature laser-triggered method to anchor $CsPbBr_3$ nanocrystals on rGO sheets without affecting the initial morphology and crystal structure of the components. The flexible and high surface area of the rGO enables the conjugation of individual metal halide perovskite nanocrystals with the emergence of new synergetic functionalities from the combination of the two counterparts. These synergetic functionalities lead to the 178- and 152-times improvement of the specific capacity of the conjugated-based electrodes compared to the single rGO and perovskite electrodes respectively.**

## 1. Introduction

To meet the world's energy needs, simultaneous development of energy conversion and energy storage systems are required. At the same time the investigation of efficient energy-assisted, low-cost, fast and solution-processable new materials capable for these energy-related applications is essential.[1,2] Metal halide perovskites hold a prominent role as active material in the photovoltaic solar cells leading to high performance, competitive to those of well-established technologies.[1,2] Nevertheless, the study of the metal halide perovskite materials in energy storage devices is still limited and mainly focused on the batteries. Hybrid organic-inorganic and all-inorganic metal halide perovskite materials in the nanocrystal[3], microcrystal[4–9] or film-like structures[10] have been proposed as anode materials in Li-based batteries with the highest specific discharge capacity to reach the value of 549 $mAh\cdot g^{-1}$ with an optimum stability of 1500 cycles.[6] On the other hand, the utilization of these materials due to their high ionic conductivity in supercapacitors as electrode elements despite focus on hybrid metal halide perovskites[11–14] , it gave the maximum capacitance of 3.32 $F\cdot cm^{-2}$ for the lead-free hybrid bismuth-halide complex: $(CN_2SH_5)_3BiI_6$ [13] and an enhanced value of 121 $F\cdot g^{-1}$ for the all-inorganic $CsPbBr_3$.[15]

In addition, the 2D graphene-related materials have been introduced as great candidates for both batteries and supercapacitors due to their high electrical conductivity.[16] They are easily fabricated and relatively cheap, with characteristic permeability, facilitating the penetration of electrolytes into the electrodes boosting the capacitance of the storage devices.[17] Their unique physicochemical features such as high morphological anisotropy, high specific surface area, different active sites, and tunable physical and chemical properties together with their flexibility and thermal/chemical stability led the scientific community to use them as anodes in Li batteries with good capacity and cycling lifetime.[18] The maximum charge capacity of graphite in Li batteries found to be up to 372 $mAh\cdot g^{-1}$, [19] while the theoretical capacity of graphene reached the value of 744 $mAh\cdot g^{-1}$. [20] Graphene flakes can buffer the volume effect of the anode materials during charge and discharge, and improve their electronic conductivity, while graphene/metal (metal oxide) composites can be used as battery cathodes to improve their cycling performance.[16,20,21] In that perspective, the improvement of supercapacitor performance can be accomplished through the fabrication of hybrid (2-components) electrodes coupling transition metal oxides with graphene-related materials taking advantage of the synergy between the electric double layer capacitors (EDLC) and pseudocapacitance originating from the different components. Specifically, transition metal oxides have been investigated as promising electrode materials for supercapacitors due to their high capacitance via quick and reversible redox processes at their surfaces, but they suffer from low conductivity.[22] Importantly, a specific capacitance value of 389 $F\cdot g^{-1}$ was reached for the hybrid $MnO_2$-graphite[23], while it was significant lower for those with rGO (135.5 $F\cdot g^{-1}$)[24] and GO (197.2 $F\cdot g^{-1}$)[25]. Significant enhanced value also found in the case of silver-doped $MnO_2$ coupled with GO (877 $F\cdot g^{-1}$)[26]. Other hybrid materials with high specific capacitance were the $Co_3O_4$ with graphene (243 $F\cdot g^{-1}$)[27] or rGO (291 $F\cdot g^{-1}$)[28], ZnO with rGO (308 $F\cdot g^{-1}$)[29], $CeO_2$ with graphite (191 $F\cdot g^{-1}$)[30], $Fe_3O_4$ with rGO (480 $F\cdot g^{-1}$)[31]. Finally, none report found related with the combination of metal halide perovskites with graphene-related materials for supercapacitors.

Very recently a new type of supercapacitors, the Zn-ion supercapacitors, has gained significant attention as one of the most promising types of electrochemical energy storage devices with the advantages of high theoretical capacity, safety, environmentally friendly nature, elements abundance (~300 times higher than lithium) and cost effectiveness.[32,33] In order to obtain high performance and long-life Zn-ion supercapacitors, new electrode materials have to be explored. 2D materials are among these elements.[33] Their low thickness can shorten the charge diffusion path during charge and discharge cycles, while their ability to endure high stress and strain without structural collapse enhances the capacity and the lifetime/stability of the energy storage devices in such type of supercapacitors.[33] Nevertheless, the electrochemical behavior of the graphene-related materials is complicated because it is correlated to the amount of the defects, functional groups and impurities.[34] Very recently, Xu et al. revealed that in Zn-ion hybrid capacitors beyond the contribution of oxygen-containing groups, an extra contribution from the reversible adsorption/desorption of $H^+$ on carbon atom of rGO sheets was confirmed.[35] Despite the great research on two-components storage materials incorporating graphene-related elements, the use of such materials for Zn-ion supercapacitors is limited. Potentially, the volume expansion and self-aggregation of the nanomaterials in the two-components materials would effectively be moderated to enhance the capacitance and prolong the lifetime of the Zn-air supercapacitors.[33] RGO has been conjugated with NbPO,[36] carbon nanotubes[37] or $Co_xNi_{2-x}P$[38] with the latter one presenting the highest specific capacitance of 356.6 $F{\cdot}g^{-1}$. In addition, there is no report to the best of our knowledge on the metal halide perovskites conjugated with 2D materials as electrode materials in Zn-ion supercapacitors.

In this work, we use for the first time the flexible high surface network of the rGO to anchor on them metal halide perovskite nanocrystals and investigate the conjugations as electrode material in Zn-ion supercapacitors. The specific capacitance for the conjugated systems was found to be extremely enhanced (107 $F{\cdot}g^{-1}$) compared to the electrodes of pure rGO (600 $mF{\cdot}g^{-1}$) and metal halide nanocrystals (700 $mF{\cdot}g^{-1}$). The rGO was functioned as a conductive channel for charge transfer and energy storage via physical adsorption and desorption on its surface (Electric double layer storage mechanism), and also as a template to induce the growth of individual metal halide perovskite nanocrystals that are responsible for the pseudocapacitance recorded for the conjugations. Furthermore, the photo-induced process using a femtosecond laser of 1026 nm wavelength was selected for the conjugation of the two materials because finely control of the nanocrystals number on the 2D material can be succeeded. In that case, the size, the morphology and the crystallinity of the nanocrystals remain unaffected from the few seconds irradiation of the mixture colloid. A procedure that overcomes the homogeneity issues rising of the anchored nanocrystals with wet chemistry methods.[39–41]

## 2. Results and discussion

Motivated from works that combine two different materials for enhancing the storage properties of the electrodes in supercapacitors[42–47] and the unique electrochemical and thermal/strain release properties of the 2D materials[18,33], conjugations of the metal halide perovskite nanocrystals with rGO flakes were fabricated using a photo-triggered method. This is the first time that such method has been utilized for the fabrication of electrode materials for Zn-based supercapacitors.

Specifically, well-crystalline and homogenous PL active $CsPbBr_3$ nanocrystals with size around 11 nm were synthesized with a room-temperature ligand-assisted re-precipitation-based method with slight modifications of the protocol previously reported by Li et al[48] (Figure S1 and S2), while before the conjugation with the amine-functionalized rGO, the exfoliation of the bulk-like flakes was required (Figure S3). The thickness of the liquid-phase exfoliated flakes was determined to be 2.3± 0.2 nm. The two materials, the purified nanocrystals (Figure 1a) and the exfoliated rGO flakes (Figure 1b), were dispersed in a common solvent (toluene) (Figure 1c) and irradiated with a femtosecond laser of 1026 nm wavelength, the set-up of which is illustrated in Figure S4. Previous work of our group revealed that the irradiation of the liquid with a femtosecond laser of 513 nm wavelength containing perovskite nanocrystals (~100nm) and GO flakes led to the conjugation of the two materials.[49] In this work, a few irradiation pulses with a femtosecond laser with 1026 nm wavelength (irradiation with lower energy) was efficient enough for the homogenous distribution of very small perovskite nanocrystals (11 nm) on the rGO flakes confirming the tuning of the anchored nanocrystals density on the rGO flakes by the number of the irradiation pulses (Figure 1d-h).

Nanocrystals attachment on the same flakes was also occurred when the two different materials were simply mixed in the common solvent due to the amine functional group on the rGO flakes (Figure 1c), with the number of the anchored perovskite nanocrystals controllably grown by the laser irradiation (Figure 1d-g). This was also confirmed by the suppression of the photoluminescence upon the laser irradiation (Figure 1i). Furthermore, the morphology and crystal structure of the perovskite nanocrystals was not altered upon the irradiation for number of pulses below $10^5$, while they were aggregated for the irradiation with 1 million pulses (Figure 1 and Figure S5). In addition, Raman spectra recorded before and after the irradiation, verified the non-alteration of the rGO component even after 60 million pulses (Figure S6).

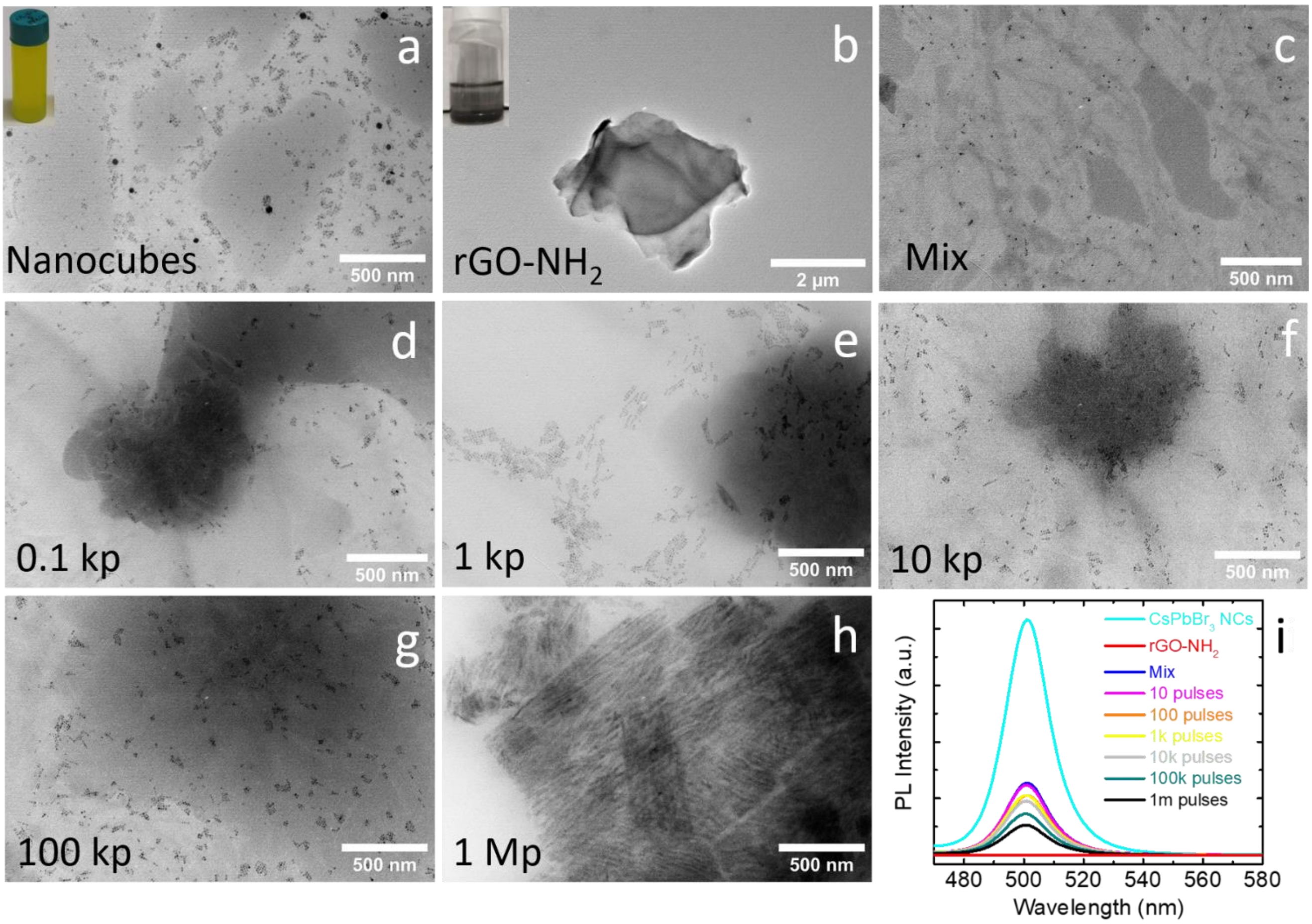

**Figure 1.** Transmission electron microscopy (TEM) images of the metal halide perovskite nanocrystals (a), the amine-functionalized rGO flakes, the mixture of the two materials before the irradiation (c), the conjugated systems after irradiation with 100 -1 million pulses (d-h), and the corresponding photoluminescence curves (i). Inset in a, b: digital photographs of the perovskite nanocrystal and rGO colloidal toluene-based dispersions.

The perovskite nanocrystal-rGO conjugations formed upon $10^5$ pulses irradiation were selected for the electrochemical characterization as electrodes for Zn-ion supercapacitors due to the large number and homogenous distribution of individually anchored nanocrystals on the flakes (Figure 1g). Specifically, the conjugates were deposited on a Ni substrate by drop casting of the solution and covered with a thin layer of $TiO_x$ deposited via Pulsed Laser Deposition (PLD). The $TiO_x$ layer as demonstrated from our previous works on metal halide nano and microcrystals as anode materials in Li-air batteries contributes significantly to the improved stability of the electrodes.[3,6]

The cyclic voltammetry (CV) curves of the conjugates-based electrodes were examined at a scan rate of 100 $mV \cdot s^{-1}$ for a potential range of -0.5 V - +1.0 V. A nearly quasi-rectangular curve is observed with a broad cathodic hump at approximately +0.65 V and an anodic hump at approximately +0.48 V, implying a pseudocapacitance behavior originating from the metal halide nanocrystals (Figure 2a). Additionally, it presents larger integral area and two orders higher magnitude of specific current than the single component electrodes (pure rGO or perovskite nanocrystals) suggesting a higher specific capacitance (i.e. the area of the CV curve is proportional to the specific capacitance) (Figure 2b and 2c).

The excellent stability of the conjugated-based electrodes during the continuous $Zn^{2+}$ intercalation/deintercalation scans was confirmed by the similar CV shapes from the first to the 100$^{th}$ scan (Figure 2a). This behavior was not in agreement with the single perovskite nanocrystal electrodes that presented deterioration in the very first cycling process (Figure 2c).

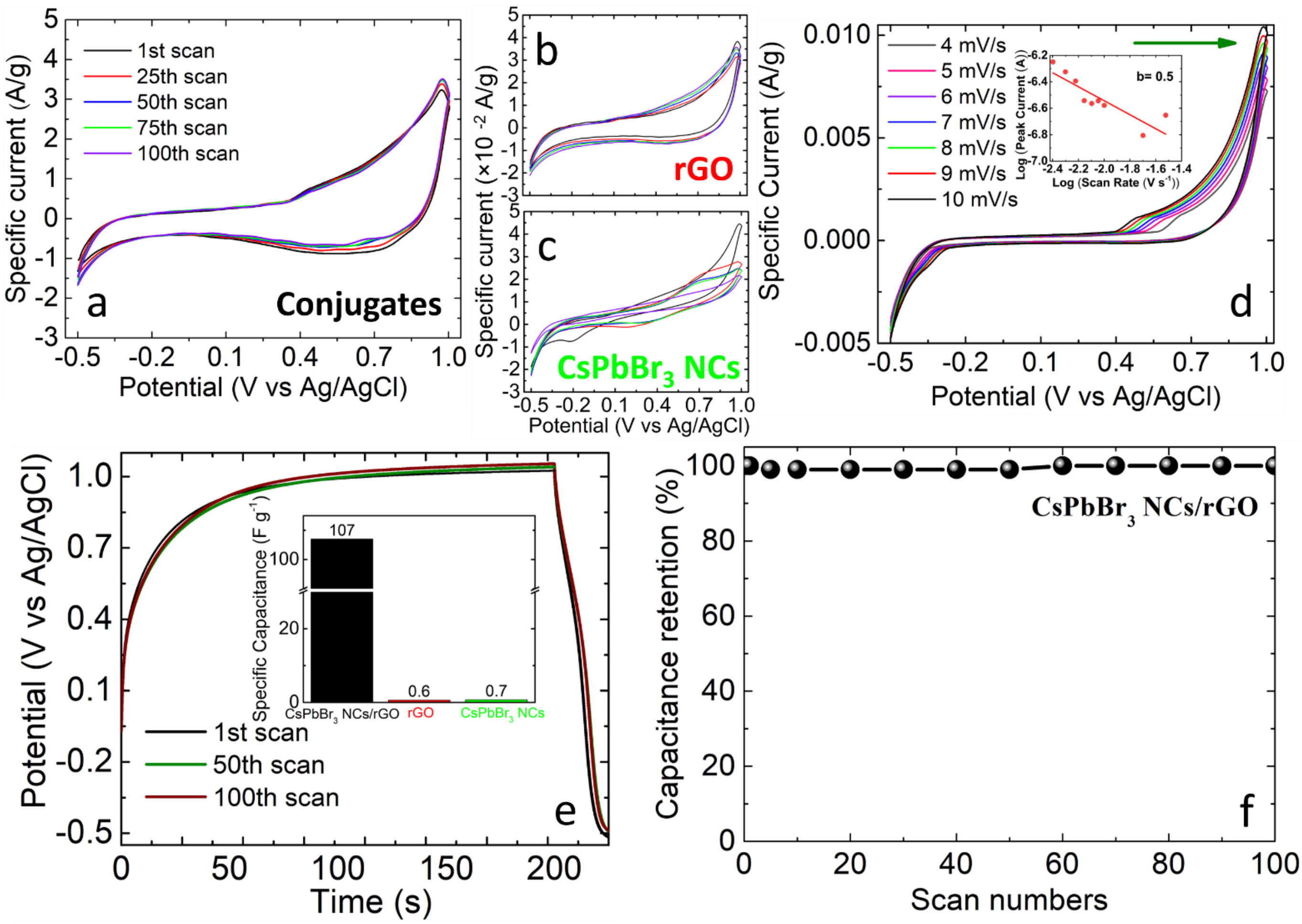

**Figure 2.** Cyclic voltammograms of the anodes based on metal halide perovskite nanocrystals/rGO conjugates (a), pure rGO (b), pure metal halide perovskite nanocrystals (c), that were traced using a 10 $mV\cdot s^{-1}$ constant scan rate and after a series of consecutive scans shown in the legends. Cyclic voltammograms for 5 to 10 $mV\cdot s^{-1}$ (d). Time dependence of the potential for 1$^{st}$, 50$^{th}$ and 100$^{th}$ scans (e). Cyclic stability of the conjugated-based electrode (f). Insets: Linear fitting for the calculation of the b value (Figure d) and comparison histogram plot for the specific capacity between the conjugation-based electrodes (black column) and single component electrodes (red and green columns for rGO and perovskite nanocrystals respectively) (Figure e).

In order to understand the mechanisms taking place during the $Zn^{2+}$ intercalation/deintercalation in the conjugated materials, CV curves were recorded at lower than 100 $mV\cdot s^{-1}$ scan rates ranging from 4 to 10 $mV\cdot s^{-1}$ (Figure 2d). As was expected, the specific current was increased with the scan rate. The peaks at ~ -0.36 V (cathodic/reduction) and ~ +0.56 V (anodic/oxidation) were shifted to more negative and positive values, respectively for higher scan rates due to the electrochemical polarization effect.[50] The polarization effect was referred to the accumulation of charge on the electrode surface for the potential range set, which can result in a shift in the potential required for the oxidation (more positive values) and reduction (more negative values) processes. It is known that the relationship between the peak current and scan rate obeys the equation $I=av^b$. The constants a and b represent the empirical parameters, i is the peak current in A and v is the scan rate in $V\cdot s^{-1}$. In particular, the b value could be utilized to evaluate the energy storage mechanism, i.e. for b=0.5 and 1.0 present a diffusion-controlled and a surface capacitive-controlled process, respectively.[51,52] By linearly fitting log(I) as a function of log(v), the b value was estimated to be 0.5 for the anodic peak (Figure 2d green arrow and inset), which implies a diffusion-controlled process. In contrary, upon increasing to 100 mV $s^{-1}$ (Figure 2a), the peaks were not detectable and the process mechanism is mainly based on pseudocapacitance indicating that the storage mechanism was an interplay between Faradaic (pseudocapacitance) and non-Faradaic processes. It is therefore a combination of Zn ions intercalation/deintercalation into the perovskite and their adsorption on the rGO' surface.

In addition, the shape of the curves in galvanostatic measurements, at an applied specific current of 0.8 $A\cdot g^{-1}$ in the potential range between -0.5 V and +1.0 V vs. Ag/AgCl, was deviated from the typical triangular shape suggesting additionally the combination of the pseudocapacitance with Faradaic capacitance[53] (Figure 2e). The estimation of specific capacitance was accomplished utilizing the equation 1 in Supplementary Information. It was found to be 107 $F\cdot g^{-1}$ for the perovskite/rGO, 600 $mF\cdot g^{-1}$ for pure $NH_2$-functionalized rGO and 700 $mF\cdot g^{-1}$ for perovskite electrodes (Figure 2e inset). The enhancement of the specific capacitance in the case of the conjugated-based electrodes can be attributed to the fact the rGO was functioned as a conductive channel for charge transfer improving the overall conductivity and acted as a flexible template to facilitate the growth of individual metal halide perovskite nanocrystals that are responsible for the pseudocapacitance. It is important to mention that there is no work with metal halide nanocrystals as psedocapacitance elements to enhance the specific capacity of rGO flakes.

It can also be observed that the potential difference is higher for the perovskite/rGO electrodes implying a more effective $Zn^{2+}$ intercalation/deintercalation process compared to that occurred in pure perovskite nanocrystal electrode (Figure S7). Hence, the particular structure allows the highest amount of charge and as a consequence the largest specific capacity. Based on the particular curves, the time constant can be calculated, which is defined as the time taken for $V_c$ (potential difference across the electrodes) to reach 63.2 % of the final value denoted as $V_s$ (i.e. this value is equal to the supply). Under these circumstances (i.e. $V_c = V_s$), the electrode is fully charged and the time constant can be found. It was estimated that the conjugates take shorter time (i.e. 13.07 s) to charge as compared with the perovskite nanocrystals (i.e. 19.74 s), while very similar with the rGO (i.e. 12.00 s), promoting the fast ions transport rate to the interior of the electrode material and resulting in enhanced efficiency.

In addition, the performance stability of the electrodes is important for energy storage applications. The conjugated-based electrode presents an excellent stability behavior remaining 100% after 100 continuous $Zn^{2+}$ intercalation/deintercalation cycles at 0.8 $A\cdot g^{-1}$ (Figure 2f). This performance is in contrast with pure perovskite electrodes where degradation was observed on the very first cycles as mentioned above.

Comparing our outcomes with other all-inorganic two-components graphene-related composites/conjugates found in the literature to be used in supercapacitors, the effective synergy of the EDLC and pseudocapacitance of the metal halide perovskites/rGO conjugates electrodes presented the highest capacitance enhancement compared to that of pure graphene-related materials (Figure 3a green arrow, Table S1). Among the systems with the highest enhancement found in the literature were the $MnO_2$ with graphene[54,55] and $MnSiO_3$ with GO[42]. It is important to notice here, that the enhancement observed for our conjugates was not directly compared with the reported materials because the electrolyte concentration utilized in the literature is mainly double and based on monovalent ions (i.e. $H^+$, $K^+$, $Na^+$) making it easier for the ions to intercalate into the framework of the material.

Unlike, the studies of all-inorganic two-components graphene-related materials as electrode materials in Zn-ion supercapacitors is limited and the capacitance of the pure graphene-related electrodes is not provided in these works in order to calculate the enhancement when the second component is conjugated. So, we decided to compare the available types of supercapacitors including their specific capacitance values (Figure 3b) to gain a better view of the literature and the novelty of our work. The capacitance of metal halide perovskite/rGO conjugates was similar to the NbPO/rGO[36], but lower to those of $Co_xNi_{2-x}P$/rGO[38], taking however into consideration that our electrolyte concentration is four times lower than the one reported for the system $Co_xNi_{2-x}P$/rGO[38]. In addition, compared to carbon nanotubes/rGO composites,[37] the capacitance of our electrodes was ~4 times higher even in 4 times lower electrolyte concentration.

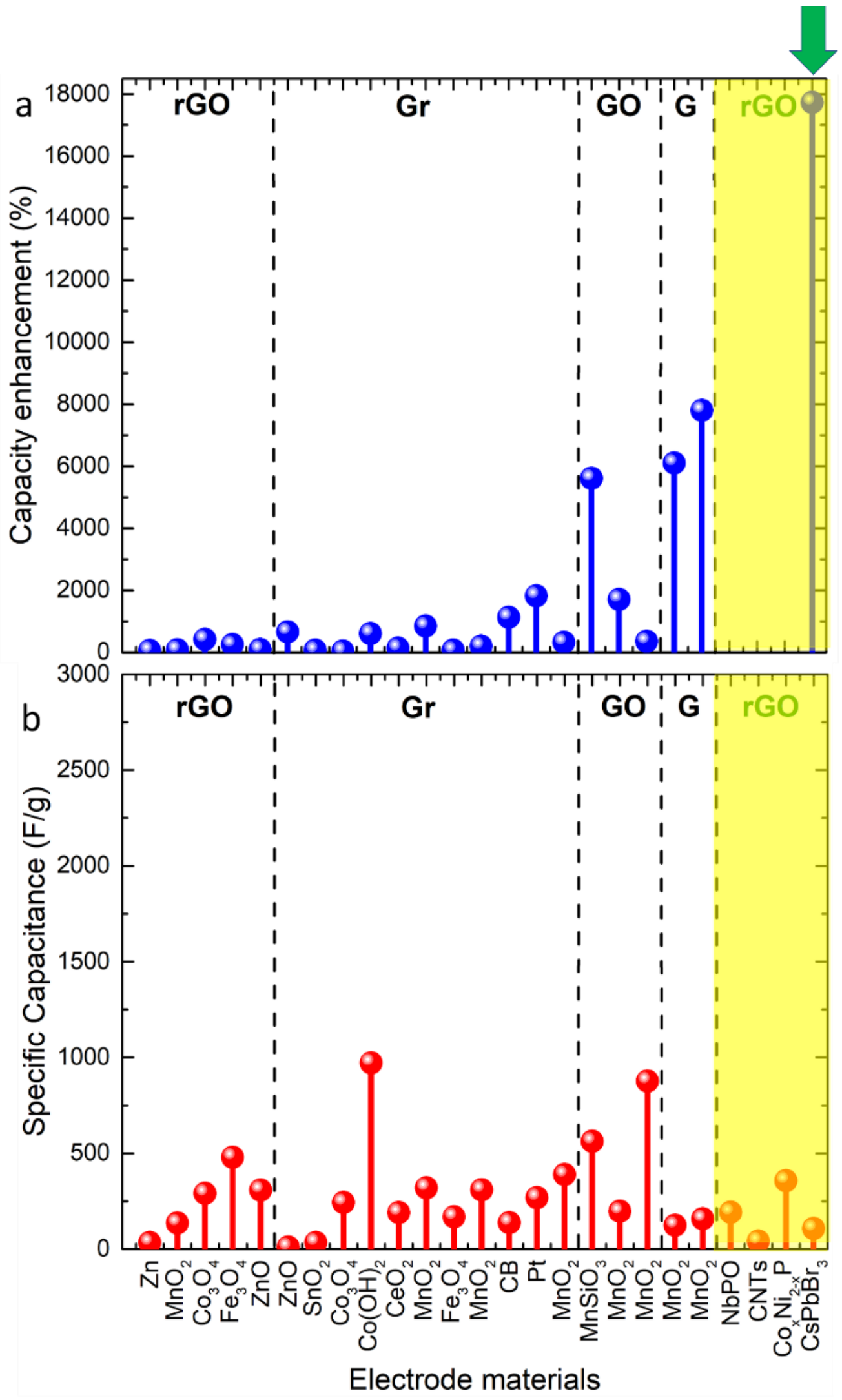

**Figure 3.** Plots depicting the comparison of the capacity enhancement (a) and their corresponding specific capacity in $F{\cdot}g^{-1}$ (b) of the $CsPbBr_3$ nanocrystals /rGO conjugates (with green arrow) with the two components composites/conjugates reported in the literature including or not a complete supercapacitor cell. The capacity value calculated compared to the value of single-2D materials electrodes. The Zn-ion supercapacitors are included in the yellow box. The capacity enhancement calculation is not possible in the Zn-ion supercapacitors due to lack of the capacity values of electrodes with the 2D materials. The materials that are presented in this figure are: Zn/rGO[56], $MnO_2$/rGO[24], $Co_3O_4$/rGO[28], $Fe_3O_4$/rGO [31], ZnO/rGO[29], ZnO/Gr[57], $SnO_2$/Gr[58], $Co_3O_4$/Gr[27], $Co(OH)_2$/Gr[59], $CeO_2$/Gr[30], $MnO_2$/Gr[60], $Fe_3O_4$/Gr, [61] $MnO_2$/Gr, [62] Carbon black (CB)/Gr[63], Pt/Gr[64], $MnO_2$/Gr[23], $MnSiO_3$/GO[42], $MnO_2$/GO[25], silver-doped $MnO_2$/GO[26], $MnO_2$-G, [55] $MnO_2$/G, [54] NbPO/rGO [36], Carbon nanotubes (CNTs)/rGO, [37] $Co_xNi_{2-x}P$/rGO[38], where rGO: reduced Graphene Oxide, Gr: Graphene, GO: Graphene Oxide, G: Graphite.

## Conclusions

It was possible to grow monodispersed conjugated metal halide perovskite nanocrystals on rGO sheets through a laser-triggered process for the precise control of the nanocrystals number on rGO. The conjugated systems presented promising performance for Zn-air supercapacitors due to the synergy of the EDLC and pseudocapacitance arising from the different components. In particular, they showed a specific capacitance value of 106 $F{\cdot}g^{-1}$ with an excellent stability remaining 100% after 100 continuous $Zn^{2+}$ intercalation/deintercalation scans. The $CsPbBr_3$/rGO electrode indicated the best stability and the highest specific capacitance as compared with the single component (i.e. rGO and $CsPbBr_3$ nanocrystals) electrodes. This is ought to rGO, which acts as a flexible conductive substrate for the growth of metal halide perovskite nanocrystals as pseudocapacitance element with an enhanced overall conductivity. Overall, we believe that the data obtained in this work is a good basis for further study of perovskite/rGO conjugates through the control of electrolyte concentration and pH value to enhance further the electrochemical performance. In addition, the room temperature laser-triggered technique selected for the conjugation of the two components provides unique opportunities for the cost-effective and large-scale synthesis of fully controllable perovskite-2D conjugates by combining nanocrystals of different morphologies and chemical phases together with multiple 2D materials in order to find the most effective combination for further enhancement of the electrochemical performance and long-term stability of the electrodes in the Zn-air supercapacitors. The amine-functionalized rGO finally found that was not the optimum choice because its EDLC capacitance was very low compared other rGO flakes. As mentioned earlier in the text, the capacity of the graphene-related materials in Zn-air supercapacitors is correlated to the amount of the defects, functional groups and impurities of them.

## Author Contributions

Conceptualization, A.K., D.V and K. B.; methodology, A.K., D.V. and K. B.; validation, N. L., S. D. and K.B.; formal analysis, A.K., D.V. and K. B investigation, A.K., D.V. and K.B.; resources, E. S., A. K.; data curation, A.K., D.V., K.B.; writing—original draft preparation, A.K. D.V.; writing—review and editing, A.K., D.V., K.B. and E.S.; visualization, A.K. and D.V.; supervision, A.K., D.V, E.S.; project administration, A.K and E.S.; funding acquisition, E.S., A. K. All authors have read and agreed to the published version of the manuscript.

## Conflicts of interest

There are no conflicts to declare.

## Acknowledgements

FLAG-ERA Joint Transnational Call 2019 for transnational research projects in synergy with the two FET Flagships Graphene Flagship & Human Brain Project - ERA-NETS 2019b (PeroGaS: MIS 5070514) is acknowledged for the financial support. We would like also to thank Dr Abdus Salam Sarkar for AFM measurements, Dr George Kenanakis for Raman experiments, Mrs Antonia Loufardaki for the film coverage of samples with PLD, Lampros Papoutsakis for the grazing incident XRD patterns and the Electron Microscopy Laboratory of the University of Crete for providing access to HRTEM and SEM facilities.

# Supporting information

**1. Characterization of the materials**

i) ***Transmission Electron Microscopy (TEM).*** TEM images were captured on a LaB6 JEOL 2100 transmission electron microscope (JEOL Ltd, Akishima, Tokyo, Japan) operating at an accelerating voltage of 200 kV. All the images were recorded by the Gatan ORIUS TM SC 1000 CCD camera (Gatan Inc. Pleasanton, CA, USA). For the TEM observation, a drop of each dispersion was placed onto a carbon-coated copper TEM grid and let to evaporate.
ii) ***Scanning Electron Microscopy (SEM).*** SEM images were collected using a JSM-6390LV instrument with tungsten filament. The samples were dried onto silicon substrates before the experiments.
iii) ***X-ray diffraction (XRD).*** Grazing incident XRD pattern was recorded using a Bruker D8 Advance XRD equipment with Cu Kα1 radiation (λ = 1.5406 Å) and a power source of 40 kV and 40 mA. The measurements were performed at RT from 10 to 50 degrees with a step of 0.02 degrees and a rate of 5 seconds/step. The XRD pattern from the perovskite nanocrystals and conjugates were collected after drop-casting their colloidal dispersion onto glass substrates and let for drying.
iv) ***UV -Vis Spectroscopy.*** The UV-Vis absorption spectra of the colloidal nanocrystal dispersion were placed in quartz cuvettes and measured using a Perkin Elmer, Lamda 950 UV/VIS/NIR spectrophotometer.
v) ***Photoluminescence (PL) Spectroscopy.*** The PL emission of the dispersions were measured at RT utilizing a Fluoromax Phosphorimeter (Horiba Ltd., Kyoto, Japan) with a 150 W Xenon continuous output ozone-free lamp. The dispersion was placed in quartz cuvettes for the measurement.
vi) ***Raman Spectroscopy.*** Raman experiments were carried out using a Nicolet Almega XR Raman spectrometer (Thermo Fisher Scientific, Waltham, MA, USA) with 473 nm laser. The samples from the dispersions were dried onto silicon substrates before the experiments whereas the rGO powder was measured directly.
vii) ***Atomic Force Microscopy (AFM).*** AFM experiments using tapping mode were employed in the Multimode Atomic Force Microscope from Digital Instruments, Bruker. The dispersions were deposited trough drop-casting onto silicon substrates and they dried before the measurements.

**2. Metal halide perovskite nanocrystals synthesis and characterization**

### i) Materials

Cesium bromide (CsBr, 99.999%), lead (II) bromide ($PbBr_2$, trace metals basis, 98%), oleic acid (technical grade, $C_{18}H_{34}O_2$, 90%), oleylamine (technical grade, $C_{18}H_{37}N$, 70%), N,N-dimethylformamide (DMF, $C_3H_7NO$, reagent, >99.9%), toluene ($C_7H_8$, ACS reagent, ≥99.5%). All chemicals were received from Sigma-Aldrich and used without any treatment.

### ii) Synthesis

$CsPbBr_3$ nanocrystals were synthesized by a precipitation-based protocol at ambient conditions reported by Li et al with slight modifications.[1] Specifically, 0.2 mmole $PbBr_2$ and 0.4 mmole CsBr were dissolved in 5 ml DMF and then 0.5 mL oleic acid (OA) and 0.25 mL oleylamine (Olam) were added in this stock solution. Following this procedure, 0.3 mL of the precursor stock solution was added dropwise in 5 ml of toluene under vigorous stirring (1200 rpm) and the solution let for stirring for 5 minutes. The color of the solution under a UV lamp is changing during the addition of the stock solution (Figure S1). The nanocrystals have been centrifuged two times to remove the excess of reactants: i) 6000 rpm for 10 minutes, the precipitated product re-dispersed in toluene and ii) 700 rpm for 7 minutes, the supernatant was collected.

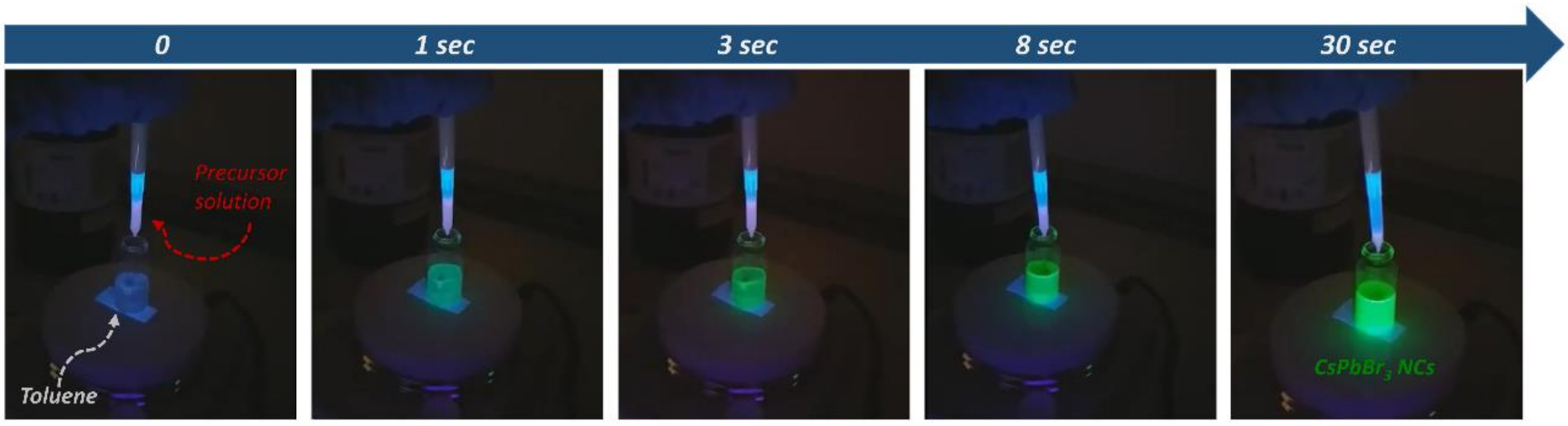

**Figure S1.** Digital photographs of the time evolution of nanocrystals formation after the addition of the precursor stock solution in the “bad” solvent (toluene) under a UV lamp.

### iii) Morphological/structural features and optical properties of the nanocrystals

The as-prepared nanocubes were quite monodispersed in size, exhibiting a narrow size distribution with edge length of 10.9 ± 2.1 nm (Figure S2a). The crystallinity of the nanocrystals was confirmed by the grazing incident XRD pattern, and the Bragg reflection peaks can be indexed to the orthorhombic $CsPbBr_3$ crystal structure (COD 4510745) (Figure S2c).

The nanocrystal solution after the centrifugation and redispersion in toluene was photoluminescent active. A single and narrow PL peak (FWHM of 17.4 nm) centered at 501 nm was observed. The cyan-green emitting PL of nanocrystal solution can be attributed to weak quantum confinement effects due to nanocrystals with dimensions below Bohr radius (7 nm).[2, 3,4] The energy gap of the $CsPbBr_3$ nanocrystals calculated to be 2.45 eV from the Tauc plot (Figure S2b).

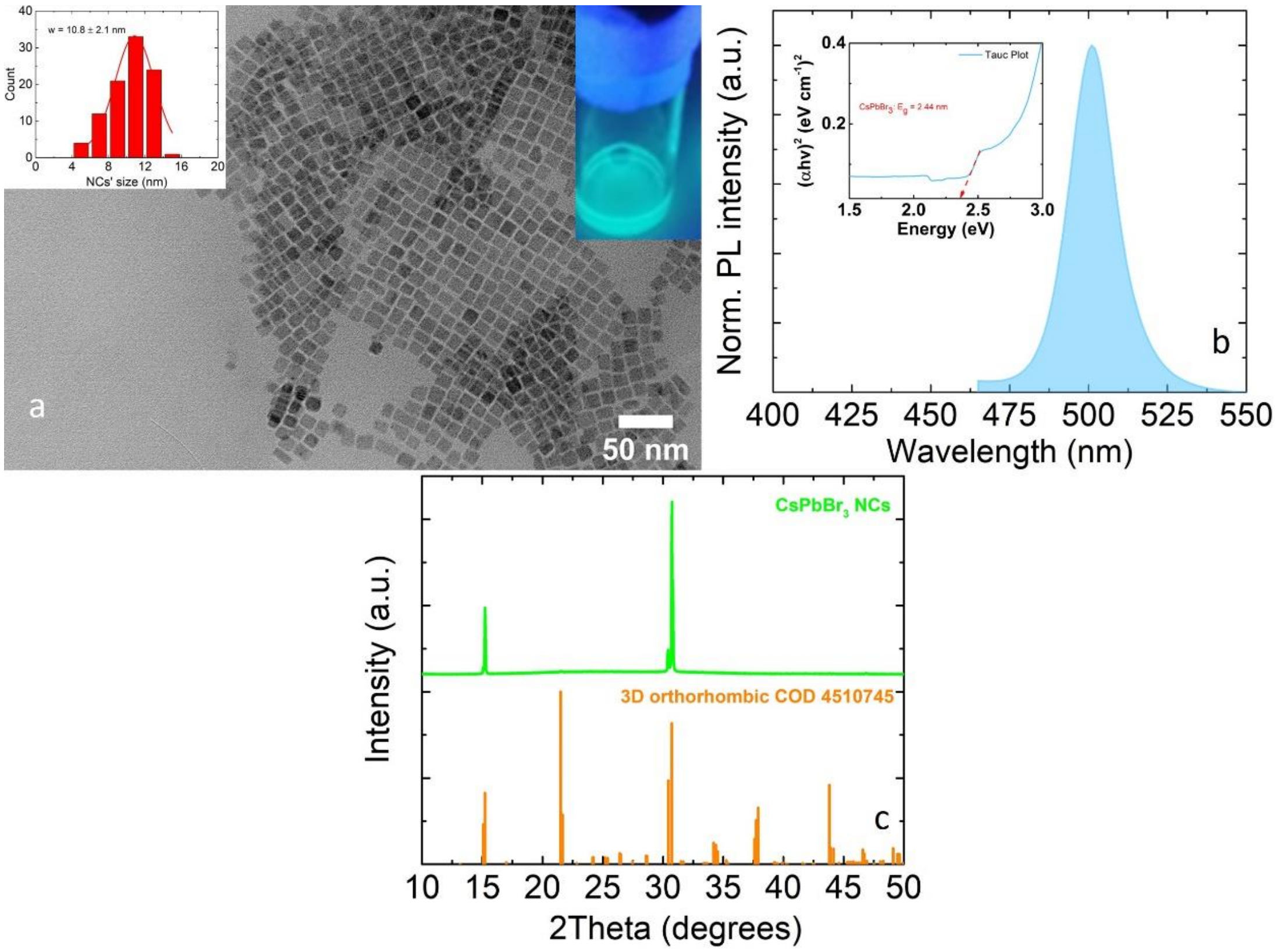

**Figure S2.** TEM image (a), Pl spectrum (b) and grazing incident XRD pattern (c) of the metal halide nanocrystals fabricated for the conjugation. Insets: size distribution diagram and photo of nanocrystal dispersion under UV lamp (a), Tauc Plot of the UV-visible absorption data for the calculation of band gap energy (b).

### 3. Reduced graphene oxide dispersion in toluene

#### i) Materials

Amine functionalized reduced graphene oxide (rGO-$NH_2$, powder, Carbon, >65 wt. %, Nitrogen, >5 wt. %), toluene ($C_7H_8$, hydrous reagent), All chemicals were received from Sigma-Aldrich.

#### ii) rGO flakes dispersion

Liquid phase exfoliation of the commercial rGO powder was employed in order to obtain rGO thinner sheets dispersed in toluene (the common solvent with the perovskite nanocrystals). Specifically, 5 mg of the amine-functionalized reduced graphene oxide was dispersed in 6 ml toluene and placed in the sonication bath for 1 hour.

#### iii) Morphological/structural features and optical properties of the nanocrystals

The exfoliation and the reduction of the flakes' size of the commercial amine-functionalized rGO were confirmed from the combination of the results from the SEM, TEM and AFM (Figure S3). Flakes with lateral sizes in the order of several microns were revealed from SEM and TEM images for the exfoliated sample while an average thickness of 2.3 ± 0.2 nm was determined from AFM. Before the exfoliation, the thickness of the rGO flakes was 7.5 ± 0.5 nm. The reduced thickness was also revealed with TEM from the different contrast between the flakes before and after the exfoliation process.

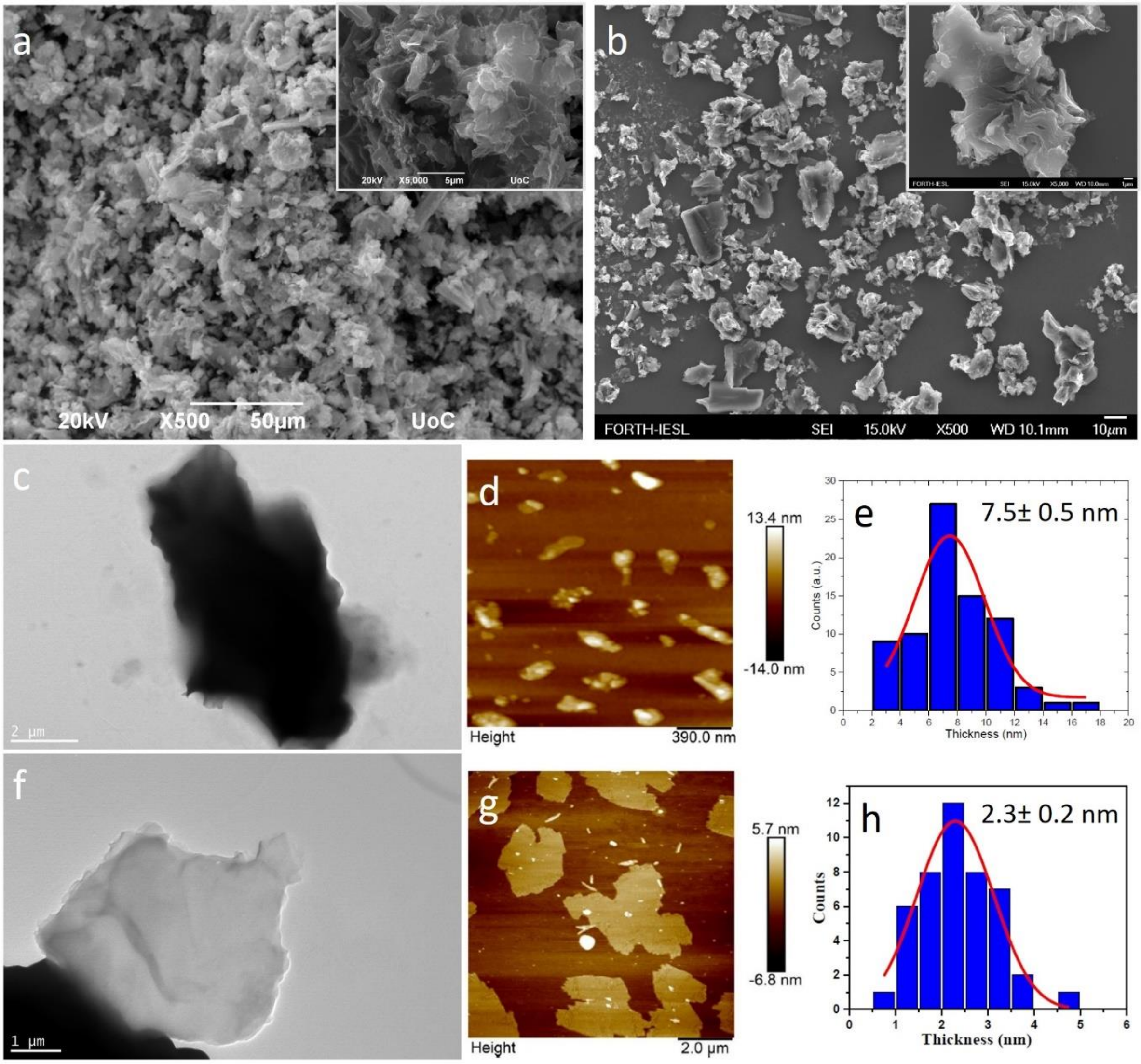

**Figure S3.** SEM (a, b), TEM (c, f) and AFM (d, g) images of the $NH_2$-functionalized rGO flakes before and after liquid-phase exfoliation. Thickness distribution diagrams of the thickness calculated from the AFM images (e, h).

**4. Photo-induced conjugation of perovskite nanocrystals on rGO flakes: Set up and irradiation conditions**

The setup used for conjugation of the two materials was described in our previous publications. [5,6] It was consisted of an Yb:KGW ultrafast pulsed laser source, two mirrors and a convex lens of 20 cm focal length (Figure S4). Contrary to our previous reports, a different laser was used for these experiments and continuous stirring was applied in order to obtain homogenous distribution of the perovskite nanocrystals on the rGO flakes. A laser of 1026 nm laser was utilized in order not to affect the small perovskite nanocrystals by the irradiation. The laser source emitted linearly polarized pulses with 170 femtoseconds at 60 kHz repetition rate at 1026 nm wavelength. The nanocrystals and the $NH_2$ functionalized rGO in the common solvent were placed in a vial and fixed 5 cm out of focus length. The Gaussian spot diameter was 700 μm at $1/e^2$, which measured and analyzed utilizing a CCD camera. The laser fluence fixed for all the irradiation experiments at 0.5 $mJ/cm^2$.

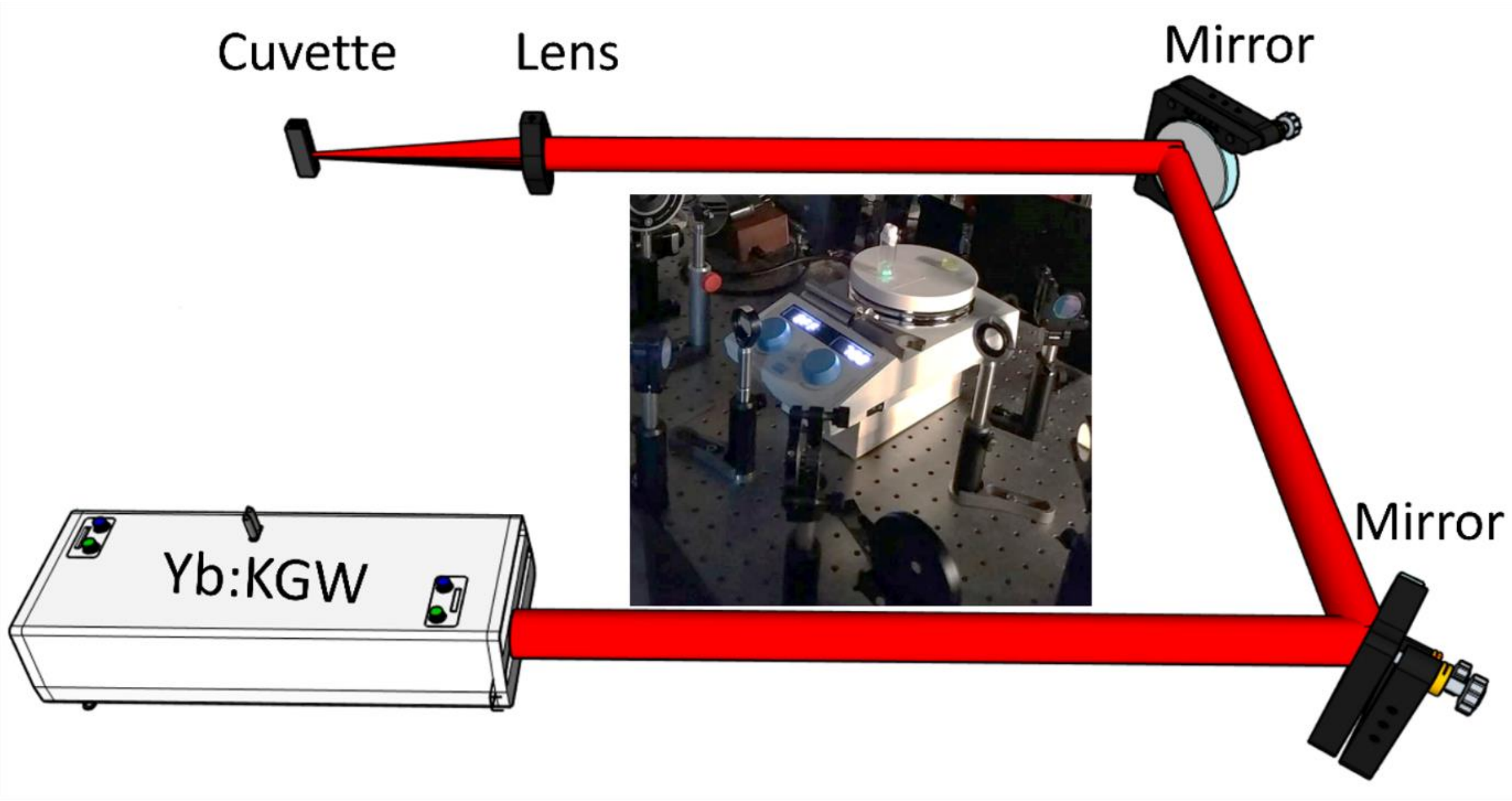

**Figure S4.** Schematic representation of the setup used for the conjugation of the metal halide perovskites with the rGO flakes utilizing a femtosecond pulsed laser of 1026 nm wavelength.

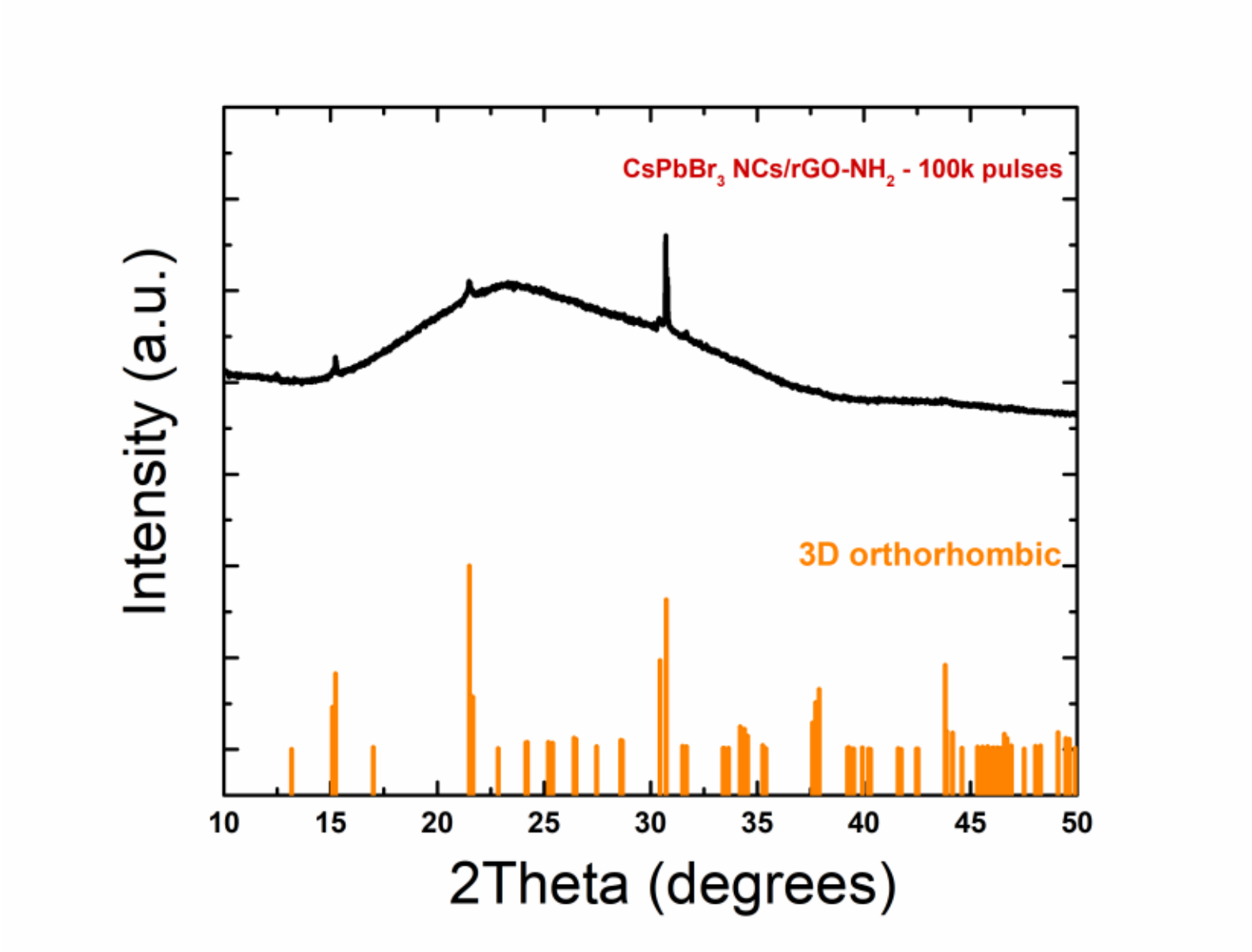

**Figure S5.** Grazing incident XRD pattern of the conjugated pattern after irradiation with $10^5$ number of pulses.

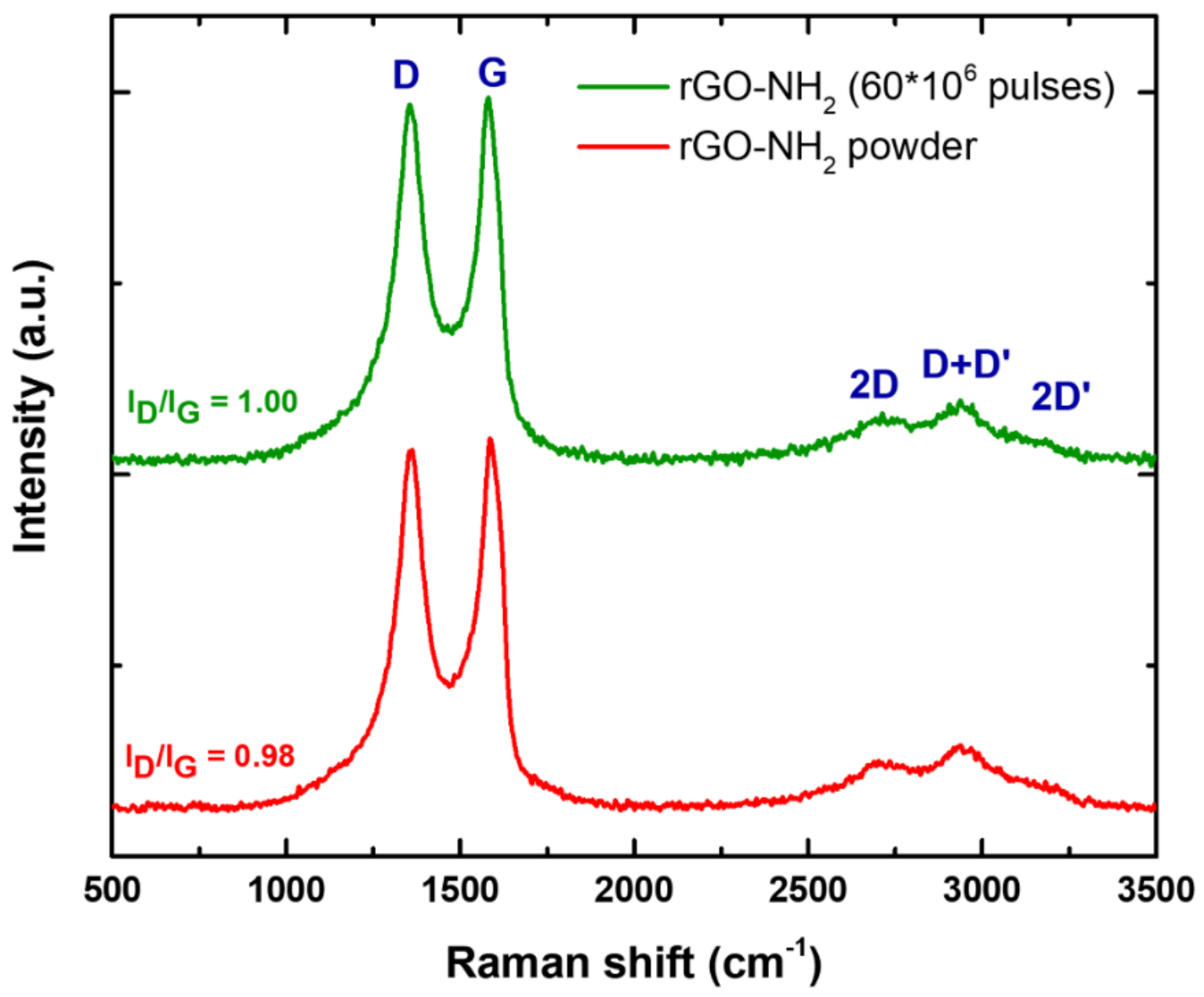

**Figure S6.** Raman spectra of the $NH_2$-functionalized rGO, before (red curve) and after laser irradiation upon 60 million pulses (green curve).

**4. Electrochemical characterization**

The metal halide perovskite nanocrystal/rGO conjugates, pure rGO and perovskite nanocrystal were deposited on Ni foil, with thickness 0.5 mm (99.98% trace metals basis) purchased from Sigma Aldrich by drop casting of the toluene-based solution and covered with a thin layer of $TiO_x$ deposited via Pulsed Laser Deposition (PLD). $TiO_2$ targets used in PLD for the coverage were prepared by pressing 1 g of 99 % pure $TiO_2$ powder purchased from Sigma Aldrich at room temperature under 4 tons for 2 min. Pulsed laser ablation of the solid targets and subsequent sample coverage were performed into a high vacuum ($3x10^{-6}$ mbar) chamber, using a KrF laser (excimer laser with 248 nm wavelength, 20 ns pulse duration and 10 Hz pulse repetition rate) emitting an energy density (fluence) of 2 $J{\cdot}cm^{-2}$. Adjusting the focusing lens, the laser spot size was optimized and an optical attenuator was used to tune the laser fluence. The target-to-substrate distance was adjusted to 5 cm and the optimum laser pulses number to 1000.

The electrochemical performance of the anode materials was evaluated in a three-electrode electrochemical cell utilizing the anode materials, Pt and Ag/AgCl as working, counter and reference electrodes, respectively. An aqueous solution of 0.5 M $ZnSO_4$ acted as an electrolyte during all measurements for a potential window of -0.5 V to +1.0 V and a scan rate of 100 $mV{\cdot}s^{-1}$. In order to understand the mechanisms taking place during the $Zn^{2+}$ intercalation/deintercalation processes, scan rates ranging from 4 $mV{\cdot}s^{-1}$ to 10 $mV{\cdot}s^{-1}$ were studied. The galvanostatic measurements were also performed at an applied specific current of 0.8 A $g^{-1}$ in the potential range between -0.5 V to +1.0 V vs. Ag/AgCl.

The specific capacitance was estimated from the equation through the galvanostatic measurements:

$C_{sp} = (I \times t)/(\Delta V \times m)$ (1)

where $C_{sp}$ is the specific capacitance by mass of the active material in $F{\cdot}g^{-1}$, I is the current applied in A, t is the charge time in s and m is the mass of the active material in g .[7]

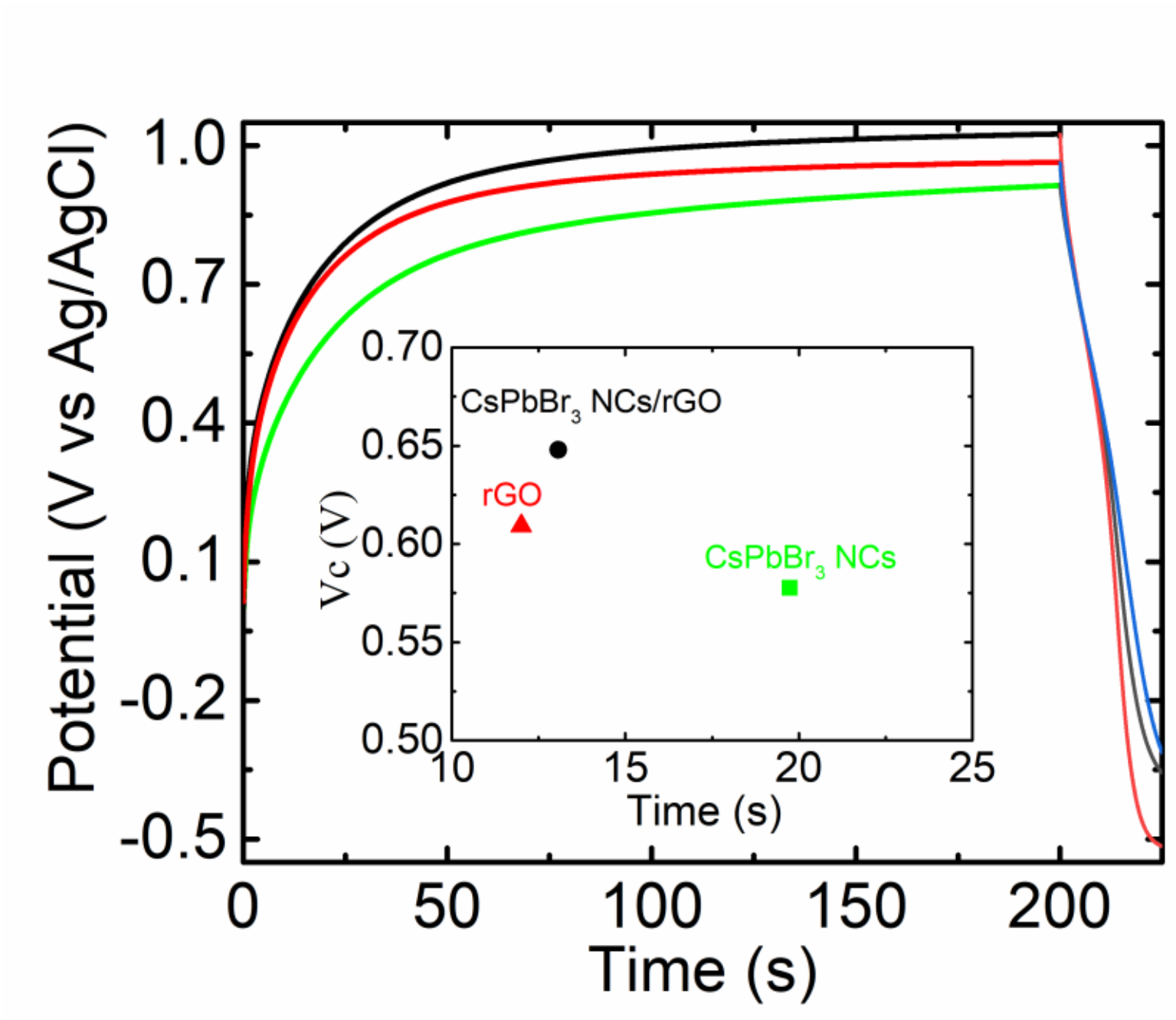

**Figure S7.** Time dependence of the potential for the conjugations (black), pure rGO (red) and pure perovskite nanocrystals electrodes.

**Table S1.** All-inorganic hybrid graphene-based composites for supercapacitors. The capacity enhancement of the hybrid/conjugated systems compared to the capacitance of the rGO used in each work has been calculated from the equation: Enhancement (%)= $[(C_{hybrid}-C_{rGO})/C_{rGO}]*100$, where $C_{hybrid}$ and $C_{rGO}$ the capacitances of the hybrid material and rGO respectively.

| **Hybrid composites** | **$C_{composite}$ (F/g)** | **$C_{2D\ material}$ (F/g)** | **Conditions** | **Enhancement (%)** | **Ref** |
|---|---|---|---|---|---|
| Zn/rGO | 33.88 | 20.78 | At 10 mV $s^{-1}$ in 1 M $H_2SO_4$ | 63 | 8 |
| $MnO_2$/rGO | 135.5 | 74.5 | At 40 mV $s^{-1}$ in1 M $Na_2SO_4$ | 81.8 | 9 |
| $Co_3O_4$/rGO | 291 | 56 | At 1 A $g^{-1}$ in 6 M KOH | 419.6 | 10 |
| $Fe_3O_4$/rGO | 480 | 139 | At 5 A $g^{-1}$ in 1 M KOH | 245.3 | 11 |
| ZnO/rGO | 308 | 157 | At 1 A $g^{-1}$ in 0.1 M $Na_2SO_4$ | 96.2 | 12 |
| ZnO/Gr | 11.3 | ~1.5 | At 10 m$Vs^{-1}$ in 1 M KCl | 653.3 | 13 |
| $SnO_2$/Gr | 34.6 | 20.7 | At 1.0 V $s^{-1}$, 1 M $H_2SO_4$ | 67.1 | 14 |
| $Co_3O_4$/Gr | 243 | 169.3 | At 10 mV $s^{-1}$ in 6 M KOH | 43.53 | 15 |
| $Co(OH)_2$/Gr | 972.5 | 137.6 | At 5 mV $s^{-1}$ in 6 M KOH | 606.8 | 16 |
| $CeO_2$/Gr | 191 | 81.57 | At 5 mV $s^{-1}$ in 3 M KOH | 134.2 | 17 |
| $MnO_2$/Gr | 320 | 34 | At 10 mV $s^{-1}$ in 1 M of $Na_2SO_4$ | 841.2 | 18 |
| $Fe_3O_4$/Gr | 169 | 101 | At 1 $Ag^{-1}$ in 1 M KOH | 67.3 | 19 |
| $MnO_2$/Gr | 310 | 104 | At 2 mV $s^{-1}$ in 1 M $Na_2SO_4$ | 198 | 20 |
| Carbon Black/Gr | 138 | 11.2 | At 10 mV $s^{-1}$ in 6 M KOH | 1132 | 21 |
| Pt/Gr | 269 | 14 | In 0.5 N $H_2SO_4$ | 1821 | 22 |
| $MnO_2$/Gr | 389 | 93 | At 1 A $g^{-1}$ in 1 M $Na_2SO_4$ | 318.3 | 23 |
| $MnSiO_3$/GO | 262.5 | 4.6 | At 0.5 A $g^{-1}$ in 1 M $Na_2SO_4$ | 5606 | 24 |
| $MnO_2$/GO | 197.2 | 10.9 | At 200 mA $g^{-1}$ in 1 M $Na_2SO_4$ | 1709 | 25 |
| Silver-doped $MnO_2$/GO | 877 | 195 | At 5 mV $s^{-1}$in 1 M of $Na_2SO_4$ | 349.7 | 26 |
| $MnO_2$/G | 124 | <2 | At 2 mV $s^{-1}$ in 1 M $Na_2SO_4$ | 6100 | 27 |
| $MnO_2$/G | 158 | <2 | At 2mAc$m^{-2}$ in 2 M $(NH_4)_2SO_4$ | 7800 | 28 |
| **Zn-ion supercapacitors** | | | | | |
| NbPO/rGO | 191.88 | | At 1 A $g^{-1}$ in 0.2 M $ZnSO_4$ | | 29 |
| Carbon nanotubes/rGO | 41.7 | | At 1 mV $s^{-1}$ in 2 M (PVA)/$Zn(CF_3SO_3)_2$ | | 30 |
| $Co_xNi_{2-x}P$/rGO | 356.6 | | At 0.5 A $g^{-1}$ in 2 M $ZnSO_4$ | | 31 |
| **Our work** | **107** | **0.600** | At 0.8 A $g^{-1}$ in 0.5 M $ZnSO_4$ | **17733** | |